# Magnetic order and 5d$^1$ multipoles in a rhenate double perovskite Ba$_2$MgReO$_6$

S. W. Lovesey[1,2] and D. D. Khalyavin[1]

[1]ISIS Facility, STFC, Didcot, Oxfordshire OX11 0QX, UK

[2]Diamond Light Source Ltd, Didcot, Oxfordshire OX11 0DE, UK

**Abstract** Structural and magnetic transitions in a double perovskite hosting 5d$^1$ Re ions are discussed on the basis of recently published high-resolution x-ray diffraction patterns [D. Hirai, *et al.*, Phys. Rev. Res. **2**, 022063(R) (2020)]. A reported structural transition below room temperature, from cubic to tetragonal symmetry, appears not to be driven by T$_{2g}$-type quadrupoles, as suggested. A magnetic motif at lower temperature is shown to be composed of two order parameters, associated with propagation vectors **k** = (0, 0, 1) and **k** = (0, 0, 0). Findings from our studies, for structural and magnetic properties of Ba$_2$MgReO$_6$, surface in predicted amplitudes for x-ray diffraction at rhenium L$_2$ and L$_3$ absorption edges, and magnetic neutron Bragg diffraction. Specifically, entanglement of anapole and spatial degrees of freedom creates a quadrupole in the neutron scattering amplitude. It would be excluded in an unexpected scenario whereby the rhenium atomic state is a manifold. Also, a chiral signature visible in resonant x-ray diffraction will be one consequence of predicted electronic quadrupole and magnetic dipole orders. A model Re wave function consistent with all current knowledge is a guide to electronic and magnetic multipoles engaged in x-ray and neutron diffraction investigations.

## I. INTRODUCTION

Electronic and magnetic properties of double perovskites hosting heavy transitions metal ions are popular topics of research. Compounds in the news include Ba$_2$MgReO$_6$ [1, 2], Ba$_2$BOsO$_6$ with B = Zn, Mg, Ca [3, 4], Ba$_2$YReO$_6$ [5, 6] and Sr$_2$MgReO$_6$ [7]. The spin-orbit coupling in the single-ion Hamiltonian increases in magnitude as $Z^4$ to a good approximation, where Z is the atomic number. For 5d-systems it is of a similar order of magnitude to both the Hund's and crystal field terms, and a range of behaviours can be realized depending on the balance between these and the magnetic exchange (the coupling constant).

The heavy transition ions possess electronic and magnetic multipoles that deflect beams of neutrons and x-rays. In addition to conventional magnetic dipoles, formed by expectation values of spin **S** and orbital **L** operators, there is evidence of quadrupoles and octupoles [1-6]. If ions occupy acentric sites, Dirac multipoles are permitted in both neutron and x-ray diffraction, e.g., a spin anapole **Ω** = (**S** × **R**) where **R** is the position operator [8, 9, 10]. A quadrupole formed from **R** and **Ω** is capable of deflecting neutrons [11].

The present study of Ba$_2$MgReO$_6$ includes proposals for additional x-ray diffraction experiments whose significance might have been overlooked, and calculations of magnetic neutron scattering amplitudes, again for future experiments. A structural phase transition from cubic to tetragonal as the sample temperature passes through ≈ 33 K, and long-range magnetic order below ≈ 18 K are outcomes of thorough studies of single crystals using conventional (Thomson) and resonant x-ray diffraction [2]. Rhenium ions occupy centrosymmetric sites in tetragonal P4$_2$/mnm that supports Re quadrupoles [2], and the evidence is that magnetic dipoles lie in the basal plane [1, 2]. Our reported studies of electronic and magnetic multipoles are based on spatial and magnetic symmetry, and atomic physics.

In the next section, we elucidate the nature of both the structural and magnetic phase transitions. Thereafter, we classify electronic and magnetic multipoles. The tetragonal paramagnetic structure admits two of five possible charge-like quadrupoles, and the family has an additional member in lower magnetic symmetry. Our findings about quadrupole orders differ from speculations that have been reported [2]. They can be tested to some extent in azimuthal-angle scans (rotation of the crystal about the reflection vector) using resonance-enhanced x-ray Bragg diffraction. A chiral signature exists as a consequence of the specific magnetic long-range order. Likely, a magnetic quadrupole formed from **R** and **Ω** is permitted, and we calculate the intensity of magnetic neutron scattering to be tested in a future experiment. All mentioned multipoles are evaluated using a simple, but wholly feasible, Re atomic wave function. Alternative estimates can be obtained from a tried and tested package for the simulation of electronic structure [12].

## II. MATERIAL PROPERTIES

The paramagnetic phase of Ba$_2$MgReO$_6$ is described by a tetragonal space-group P4$_2$/mnm (No. 136, crystal class 4/mmm) with Re ions at sites 2a (m.mm site symmetry) [2]. Cell parameters a = b ≈ 5.7136 Å and c ≈ 8.0849 Å. Lattice vectors and origin of the subgroup in respect of the parent cubic structure Fm$\bar{3}$m (No. 225) are (½, −½, 0), (½, ½, 0), (0, 0, 1) and (0, 0, 0), respectively. Associated Miller indices *h*, *k*, *l* are integers. The propagation vector for the symmetry lowering from cubic to tetragonal is **k** = (0, 0, 1), so even if we use the parent cubic cell for indexation then some of the indices will violate the F-centring condition (but they still will be integers).

Magnetic long-range order is a combination of two order parameters, an antiferro (AFM) associated with **k** = (0, 0, 1) and a ferro (FM) associated with **k** = (0, 0, 0). The two orders are depicted in Fig. 1, and their individual symmetries are [13]:
(AFM) P$_I$nnm (BNS setting, No. 58.404), basis = {(0, 0, 1),(½, ½, 0),(−½, ½, 0)}, origin at (0, 0, 0), Re use sites 2a.
(FM) Im′m′m (No. 71.536), basis = {(0, 0, −1),(−½ ,½ ,0),(½, ½, 0)}, origin at (0, 0, 0), Re use sites 2a.
The presence of both order parameters implies that the symmetry of the magnetic phase is
Pnn′m′ (No. 58.398, magnetic crystal-class mm′m′), using a basis **ξ** = (½, ½, 0), **η** = (−½, ½, 0), **ζ** = (0, 0, 1), origin (0, 0, 0), and Re in sites 2a with symmetry 2′/m′. A spontaneous magnetization is permitted in the magnetic crystal-class. Also, a nonlinear magnetoelectric effect due to an invariant of the type HEE alone. Such symmetry is permitted for rare-earth garnets and a nonlinear effect has been observed.

An interesting point is that the magnetic space-group allows the structural distortions with the P4$_2$/mnm symmetry. This means that there must be a coupling between the AFM, FM and these structural distortions. Specifically, there is a trilinear free-energy invariant which is just a product of three order parameters ∝ {σ · λ · μ}, with σ-structural distortions, λ-AFM and μ-FM order-parameters. This energy term indicates that in the presence of the P4$_2$/mnm structural distortions, a condensation of AFM order will induce FM component and vice versa (FM will necessarily induce AFM). This makes the scenario to be symmetry consistent.

Regarding symmetry of the quadrupoles in the tetragonal phase, the subduction frequency for the 2a site symmetry irreps in the global distortions associated with the X2+ representation reducing the symmetry down to P4$_2$/mnm. The two-dimensional E$_g$ site irrep has a non-zero subduction frequency, while the subduction frequency of the three-dimensional T$_{2g}$ site irrep is zero. Specifically, T$_{2g}$-type quadrupoles, defined in the cubic (a, b, c) basis (Fm$\bar{3}$m) depicted in Fig. 1, cannot drive the transition to P4$_2$/mnm. In fact, T$_{2g}$ distortions are secondary and associated with **k** = (0, 0, 0), meaning some ferro ordering of the T$_{2g}$-type quadrupoles should be allowed but they cannot drive the structural transition.

### III. MULTIPOLES

Rhenium ions occupy sites with inversion symmetry and all multipoles $\langle t^K_Q \rangle$, with rank K and projections $-K \leq Q \leq K$, are axial (parity-even). The complex conjugate of our multipoles satisfy $\langle t^K_{-Q} \rangle = (-1)^Q \langle t^K_Q \rangle^*$ and the diagonal $\langle t^K_0 \rangle$ is purely real, with a phase convention $\langle t^K_Q \rangle = [\langle t^K_Q \rangle' + i\langle t^K_Q \rangle'']$ for real and imaginary parts labelled by single and double primes, respectively. Cartesian and spherical components of a dipole **R** = (x, y, z) are related by x = (R$_{-1}$ − R$_{+1}$)/√2, y = i(R$_{-1}$ + R$_{+1}$)/√2, z = R$_0$. For the paramagnetic state, contiguous to long-range magnetic order in the phase diagram, we use P4$_2$/mnm (No. 136), for which Q = 2n and $(-1)^n \langle t^K_{-Q} \rangle = \langle t^K_Q \rangle$. Rank K is even for charge-like multipoles viewed in x-ray diffraction enhanced by E1-E1 (2p ↔ 3d) or E2-E2 (1s, 2s ↔ 3d) absorption events, for which the time signature σ$_θ$ = $(-1)^K$. Applied to quadrupoles, K = 2, projections Q = 0 and ±2. Site symmetry permits the diagonal quadrupole $\langle t^2_0 \rangle$ and one off-diagonal component that is purely imaginary, $\langle t^2_{+2} \rangle = -\langle t^2_{+2} \rangle^*$, i.e., the quadrupole has ξη-like angular symmetry. In the magnetic phase, sites 2a in Pnn'm' demand Q + K = 2n. Thus, Q = 0 and ±2 for K = 2, and there is no additional constraint from site symmetry on the off-diagonal $\langle t^2_{\pm 2} \rangle$.

A rhenium atomic wave function, based on the configuration 5d$^1$ (S = ½, L = 2), has two-fold rotation symmetry about the crystal c-axis, and yields a magnetic dipole confined to the ξ-η plane, as in Fig. 1. Quadrupoles in the paramagnetic phase impose an additional constraint. Candidate wave functions are linear combinations of d-orbitals d$_m$, with projections −2 ≤ m ≤ 2, that we define with respect to a rhombically distorted octahedron with axes parallel to lines joining the central Re ion to each pair of O ligand ions. Suitable combinations of orbitals are d$_0$ and d$_{\pm 2}$, or d$_{+1}$ and d$_{-1}$. We elect to use a feasible minimal model, for more than illustrative purposes, defined by |u⟩ = [αd$_0$ + βd$_{+2}$] with α purely real and β complex, and α$^2$ + |β|$^2$ = 1 for normalization. We shall find that orbital angular momentum is quenched, and α is fixed by the spin moment in the ξ-η plane. Off-diagonal, charge-like quadrupoles arise from an admixture of d orbitals, while $[\alpha^2 - |\beta|^2]$ measures the total strength of the diagonal quadrupole.

Saturation values of magnetic multipoles are calculated with a normalized ground state $|g\rangle = [|u\rangle + \exp(i\phi)|\hat{u}\rangle]/\sqrt{2}$, where $|\hat{u}\rangle$ is the conjugate component of the Kramers doublet, and the angle $\phi$ specifies orientation of dipoles in the $\xi$-$\eta$ plane. Composite spin-orbital states required in $|u\rangle$ are best represented by total angular momenta using $j = 3/2$, $j' = 5/2$, and two projections $m = 1/2$, $m' = 5/2$ (S-L coupling scheme),

$$d_0\uparrow = [\sqrt{2}\,|jm\rangle + \sqrt{3}\,|j'm\rangle]/\sqrt{5}, \quad d_{+2}\uparrow = |j'm'\rangle. \tag{1}$$

Multipoles $\langle t^K{}_Q\rangle$ in orthogonal coordinates ($\xi$, $\eta$, $\zeta$) for the tetragonal cell are related to $\langle T^K{}_Q\rangle = \langle g|T^K{}_Q|g\rangle$ by a rotation about the $\zeta$-axis through 45°, namely, $\langle t^K{}_Q\rangle = \exp(iQ\pi/4)\,\langle T^K{}_Q\rangle$. The construction of $|g\rangle$ ensures the requirement $\langle t^1{}_0\rangle = 0$ is fulfilled. In the paramagnetic phase, $\langle T^2{}_{+2}\rangle = [\alpha\beta\,\langle d_{+2}|T^2{}_{+2}|d_0\rangle]$ is purely real for real coefficients, whereupon $\langle t^2{}_{+2}\rangle = -\langle t^2{}_{-2}\rangle$ as required by site symmetry. The diagonal quadrupole $\langle t^2{}_0\rangle = [\alpha^2\langle d_0|T^2{}_Q|d_0\rangle + |\beta|^2\langle d_{+2}|T^2{}_0|d_{+2}\rangle]$. Turning to the magnetic phase, $\langle \mathbf{L}\rangle = 0$, and spin components are $\langle S_\xi\rangle = [\alpha^2 \cos(45° + \phi)/2]$, $\langle S_\eta\rangle = [\alpha^2 \sin(45° + \phi)/2]$, $\langle S_\zeta\rangle = 0$. An observed magnetic moment $\approx 0.3\ \mu_B$ implies $\alpha^2 \approx 0.3$ and $\alpha|\beta| \approx \pm 0.46$ [1].

Multipoles observed in resonance enhanced x-ray diffraction have different values at different absorption edges. Dependence on the total angular momentum of the core states, 1/2 for $L_2$ and 3/2 for $L_3$, is carried by reduced matrix elements that obey sum-rules [8, 14]. For $L_2$ and $L_3$ edges $\langle t^1\rangle_{L3} + \langle t^1\rangle_{L2} = -\langle \mathbf{L}\rangle_d/(10\sqrt{2})$, and $\langle t^2\rangle_{L3} + \langle t^2\rangle_{L2} = \langle\{\mathbf{L}\otimes\mathbf{L}\}^2\rangle_d/30$, with $\{\mathbf{L}\otimes\mathbf{L}\}^2{}_0 = [3(L_\zeta)^2 - L(L+1)]/\sqrt{6}$ for the diagonal element of the tensor product. The dipole $\langle T^1{}_{+1}\rangle = [\exp(i\phi)\,\langle u|T^1{}_{+1}|\hat{u}\rangle/2]$, and the matrix element therein for $L_2$ is denied contributions from $j'$ by the dipole selection rule. On the other hand, there are contributions to the matrix element from both atomic states, $j$ and $j'$, at $L_3$. Projections $Q = 1, 3, 5$ are allowed in the matrix element $\langle u|T^K{}_Q|\hat{u}\rangle$, which contribute to magnetic octupoles ($K = 3$) and triakontadipoles ($K = 5$).

In summary, our model Re atomic wave function yields the following guides to multipoles observed in diffraction enhanced by an E1-E1 event,

$L_2$ edge; $\langle T^1{}_{+1}\rangle = -\exp(i\phi)\,(1/45)\,\alpha^2$, $\langle T^2{}_0\rangle = -(15\sqrt{6})^{-1}\,\alpha^2$, $\langle T^2{}_{+2}\rangle = 0$. (2)

$L_3$ edge; $\langle T^1{}_{+1}\rangle = \exp(i\phi)\,(1/45)\,\alpha^2$, $\langle T^2{}_0\rangle = (15\sqrt{6})^{-1}\,[3 - 5\alpha^2]$, $\langle T^2{}_{+2}\rangle = (5\sqrt{6})^{-1}\,\alpha\beta^*$.

Recall that, $\beta$ is set purely real in the paramagnetic phase. Dipole results fit with quenched orbital angular momentum and the aforementioned dipole sum-rule. A successful analysis similar to the one proposed here for strontium iridate uses a ground state for which $\langle \mathbf{t}^1\rangle_{L2} = 0$ [15]. Evidently, experimental results for diffraction at the $L_2$ edge of Re in $Ba_2MgReO_6$ would be a valuable asset in determining the atomic ground state.

## IV. RESONANT X-RAY DIFFRACTION

The photon scattering length is developed in the small quantity $E/mc^2$, where E is the primary energy ($mc^2 \approx 0.511$ MeV). At the second level of smallness in this quantity it contains resonant processes that may dominate all other contributions should E match an atomic resonance $\Delta$. Assuming that virtual intermediate states are spherically symmetric, to a good approximation, the scattering length $\approx \{F_{\nu\mu}/(E - \Delta + i\Gamma/2)\}$ in the region of the resonance, where $\Gamma$ is the total width of the resonance. The numerator $F_{\nu\mu}$ is a dimensionless amplitude, or unit-cell structure factor, for Bragg diffraction in the scattering channel with primary (secondary) polarization $\mu$ ($\nu$). Henceforth, we use ($\nu\mu$) in place of $F_{\nu\mu}$ to abbreviate notation. By convention, $\sigma$ labels polarization normal to the plane of scattering, and $\pi$ denotes polarization within the plane. In the nominal setting, axis (x, y, z) in which $\sigma$-polarization is parallel to the z-axis and the reflection vector is parallel to $-$ x, coincide with tetragonal cell edges labelled ($\xi$, $\eta$, $\zeta$). The Bragg angle $\theta$ for a photon energy $E \approx 10.535$ eV is calculated from $\sin(\theta) \approx 0.1030 \sqrt{[h^2 + k^2 + l^2/2]}$.

A structure factor for diffraction is $\Psi^K_Q = [\exp(i\boldsymbol{\kappa} \cdot \mathbf{d}) \langle O^K_Q \rangle_{\mathbf{d}}]$, where the Bragg wave vector $\boldsymbol{\kappa}$ is defined by integer Miller indices ($h$, $k$, $l$), and the implied sum in $\Psi^K_Q$ is over all Re sites $\mathbf{d}$ in a unit cell of the tetragonal structure. For the paramagnetic and magnetic structures of interest, P4$_2$/mnm and Pnn′m′, we obtain a generic result,

$$\Psi^K_Q = [\langle O^K_Q \rangle + (-1)^{h+k+l}(-1)^K \langle O^K_{-Q} \rangle]. \qquad (3)$$

Different site symmetries for Re ions alone distinguish the paramagnetic and magnetic structures. Basis allowed reflections (K even and Q = 0) are indexed by $h + k + l$ even.

Scattering amplitudes for space-group forbidden (0, 0, $l$) with $l$ odd are [8, 16],

$$(\sigma'\sigma) = 2 \sin(2\psi) \langle t^2_{+2} \rangle'', \ (\pi'\pi) = \sin^2(\theta) (\sigma'\sigma), \qquad (4)$$

$$(\pi'\sigma) = -(\sigma'\pi) = 2[-i \cos(\theta) \cos(\psi) \langle t^1_{+1} \rangle'' + \sin(\theta) \cos(2\psi) \langle t^2_{+2} \rangle''].$$

In these expressions, $\psi$ is the angle of rotation about the reflection vector, and the $\xi$-axis is normal to the plane of scattering at $\psi = 0$. Intensity in the rotated channel of polarization, say, is proportional to $|(\pi'\sigma)|^2$, and it includes a magnetic dipole. The latter, $\langle t^1_{+1} \rangle''$, is allowed different from zero in the magnetic phase, below $\approx 18$ K, where it is proportional to the $\eta$ component of the magnetic dipole. Diffraction in the paramagnetic phase, bracketed by temperatures $\approx 18$ K and $\approx 33$ K [1, 2], is usually referred to as Templeton-Templeton scattering, and it is solely created by the quadrupole $\langle t^2_{+2} \rangle''$ with $\xi\eta$-like angular symmetry.

Scattered intensity picked out by circular polarization in the primary photon beam = $P_2 \Upsilon$ with [17, 18],

$$\Upsilon = \{(\sigma'\pi)^*(\sigma'\sigma) + (\pi'\pi)^*(\pi'\sigma)\}'', \qquad (5)$$

and the Stokes parameter $P_2$ (a purely real pseudoscalar) measures helicity in the primary x-ray beam. Since intensity is a scalar quantity, $\Upsilon$ and $P_2$ possess identical discrete symmetries, specifically, both scalars are time-even and parity-odd (polar). Partial intensity $\Upsilon$ different from zero is a signature of a chiral motif of electronic and magnetic multipoles, of course. From results in Eqs. (4),

$$\Upsilon(0, 0, l) = -2 \langle t^1_{+1} \rangle'' \cos(\theta) \cos(\psi) (\sigma'\sigma) [1 + \sin^2(\theta)]. \qquad (6)$$

Notably, $\Upsilon(0, 0, l)$ is an odd function of the azimuthal angle, and a non-zero value is specific to long-range magnetic order permitted by Pnn'm'.

Next, space-group forbidden $(h, k, 0)$ with $h + k$ odd. Let $A^1_1 = 2i \cos(\delta) \langle t^1_{+1}\rangle''$, $B^1_1 = -2 \sin(\delta) \langle t^1_{+1}\rangle''$, $A^2_2 = -2 \sin(2\delta) \langle t^2_{+2}\rangle''$, $B^2_2 = 2i \cos(2\delta) \langle t^2_{+2}\rangle''$, where $\delta$ is the angle subtended by the reflection vector and $-x$, e.g., $\cos(2\delta) = (h^2 - k^2)/(h^2 + k^2)$. Amplitudes include,

$$(\sigma'\sigma) = -\sin^2(\psi) A^2_2, \tag{7}$$

$$(\pi'\sigma) = \cos(\theta) \cos(\psi) A^1_1 + i \sin(\theta) B^1_1 - (1/2) \sin(\theta) \sin(2\psi) A^2_2 + i \cos(\theta) \sin(\psi) B^2_2.$$

The chiral signature,

$$\Upsilon(h, k, 0) = A^2_2 \{\cos(\theta) \cos(\psi) [1 - (1 + \sin^2(\theta)) \sin^2(\psi)] A^1_1{''} \tag{8}$$

$$+ \sin(\theta) [1 + \cos^2(\theta) \sin^2(\psi)] B^1_1\},$$

is proportional to $[\langle t^1_\eta\rangle \langle t^2_{+2}\rangle'']$, and it is the same as $\Upsilon(0, 0, l)$ in this respect. However, the two signals differ as functions of the azimuthal angle, with $\Upsilon(h, k, 0)$ an even function of $\psi$. Chiral signatures, Eqs. (6) and (8), according to our model Re wave function are zero at the $L_2$ edge since it predicts $\langle t^2_{+2}\rangle_{L2} = i\langle T^2_{+2}\rangle_{L2} = 0$.

## V. MAGNETIC NEUTRON DIFFRACTION

All multipoles are time-odd (magnetic), and rank $K = 1, 2, ... 5$ for a d-state. Even rank multipoles arise from a mixture of $j = 3/2$ and $j' = 5/2$ in Eq. (1) alone [9, 11]. The present calculation retains dipoles and quadrupoles in the intermediate scattering amplitude $\langle Q\rangle$ for Bragg diffraction. Multipoles for neutron scattering $\langle\tau^K\rangle$ have a strong dependence on the magnitude of the scattering wave vector imparted through spherical Bessel functions $\langle j_n(\kappa)\rangle$ averaged over a radial wave function. The dipole $\langle\tau^1\rangle$ is a sum of $\langle j_0(\kappa)\rangle$ and $\langle j_2(\kappa)\rangle$, and $\langle\tau^2\rangle \propto \langle j_2(\kappa)\rangle$. By definition, $\langle j_0(0)\rangle = 1$ while $\langle j_n(0)\rangle = 0$ for $n \geq 2$, and $\langle j_2(\kappa)\rangle$ is a maximum near $\kappa \approx 3.8$ Å$^{-1}$ just short of the first zero in $\langle j_0(\kappa)\rangle$, cf. Fig. 2 [19].

Intensity of a Bragg spot $= |\langle Q_\perp\rangle|^2$, where the operator $Q_\perp = \{\kappa^{-2} [\kappa \times (Q \times \kappa)]\}$ with,

$$Q = \exp(iR_j \cdot \kappa) [S_j - \kappa^{-2} (i/\hbar)(\kappa \times p_j)], \tag{9}$$

and the implied sum is over all unpaired electrons. In Eq. (9), **R** and **p** are conjugate operators for electronic position and linear momentum, respectively. Even rank multipoles are created by the first, spin-dependent contribution, and they represent entanglement of **R** and the spin anapole $\Omega$ [9, 11]. Amplitudes for basis forbidden reflections, $h + k + l$ odd, depend on two purely real combinations of multipoles $A = -i[\langle\tau^1_{+1}\rangle + \langle\tau^1_{-1}\rangle] = -\sqrt{2} \langle\tau^1_\eta\rangle$, and $B = [\langle\tau^2_{+1}\rangle - \langle\tau^2_{-1}\rangle] = 2\langle\tau^2_{+1}\rangle'$ with $\xi\zeta$-like angular symmetry. In terms of a unit vector $(p, q, r) = \kappa/\kappa$,

$$\langle Q_\xi\rangle \approx -p q\sqrt{3} B, \quad \langle Q_\eta\rangle \approx -(3/\sqrt{2}) A + (p^2 - r^2)\sqrt{3} B, \quad \langle Q_\zeta\rangle \approx q r \sqrt{3} B. \tag{10}$$

The purely magnetic contribution to a Bragg diffraction pattern can be recognized by a non-zero spin-flip intensity [5, 20, 21],

$$SF = |\langle \mathbf{Q}_\perp \rangle - \mathcal{P}(\mathcal{P} \cdot \langle \mathbf{Q}_\perp \rangle)|^2, \tag{11}$$

where $\mathcal{P} = (\mathcal{P}_\xi, \mathcal{P}_\eta, \mathcal{P}_\zeta)$ is a unit vector in the direction of the polarization in the primary neutron beam. For $\mathcal{P}$ parallel to the Bragg wave vector $\kappa$ the result is $SF = |\langle \mathbf{Q}_\perp \rangle|^2$. We shall consider the opposite extreme $\mathcal{P} \perp \kappa$ [21].

Intensities of Bragg spots indexed by (0, 0, $l$) with $l$ odd are $|\langle \mathbf{Q}_\perp \rangle|^2 = \langle Q_\eta \rangle^2$, while $SF = [\mathcal{P}_\xi \langle Q_\eta \rangle]^2$ for $\mathcal{P}_\zeta = 0$, and $\langle Q_\eta \rangle$ in Eq. (10) is a simple sum of the dipole and quadrupole. For the second class of basis forbidden Bragg spots $\kappa = \kappa(p, q, 0)$ we obtain,

$$|\langle \mathbf{Q}_\perp \rangle|^2 = p^2 \{\sqrt{3}B - 3A/\sqrt{2}\}^2. \tag{12}$$

Intensity is zero for $\kappa$ parallel to the crystal $\eta$-axis. Since $\langle Q_\zeta \rangle = 0$ for r = 0 one has $\mathcal{P} \cdot \langle \mathbf{Q}_\perp \rangle = 0$ and $SF = |\langle \mathbf{Q}_\perp \rangle|^2$ for $\mathcal{P} = (0, 0, 1)$.

Our minimal model of the Re atomic wave function discussed in Section III yields estimates,

$$A = -(2\sqrt{2}/3) \langle S_\eta \rangle [\langle j_0(\kappa) \rangle - (1/7) \langle j_2(\kappa) \rangle], \quad B = (10/(7\sqrt{3})) \langle S_\eta \rangle \langle j_2(\kappa) \rangle. \tag{13}$$

In this case, the dipole $A \propto \langle \tau^1_\eta \rangle$ and quadrupole $B \propto \langle \tau^2_{+1} \rangle'$ are proportional to the $\eta$-component of the magnetic moment. Radial integrals $[\langle j_0(\kappa) \rangle - (1/7) \langle j_2(\kappa) \rangle]$ and $\langle j_2(\kappa) \rangle$ appearing in A and B are displayed in Fig. 2 [19]. Note that, use of A in Eq. (12) returns the standard result whereby $|\langle \mathbf{Q}_\perp \rangle|^2$ is proportional to the square of the magnetic moment in the forward direction of scattering, where $\langle j_2(\kappa) \rangle \approx 0$.

## VI. DISCUSSION AND CONCLUSIONS

In summary, we have studied consequences of a cubic-to-tetragonal structural transition and magnetic long-range order reported at temperatures ≈ 33 K and ≈ 18 K, respectively, in $Ba_2MgReO_6$ [1, 2]. Our magnetic motif, illustrated in Fig. 1, is composed of ferro- and antiferromagnetic orders. Moreover, the corresponding space-group Pnn′m′ (No. 58.398) allows the structural distortions with the tetragonal $P4_2$/mnm symmetry, i.e., there must be a coupling between the two magnetic orders and structural distortion. Rhenium site symmetry remains centrosymmetric, and Dirac multipoles are thereby forbidden. The magnetic space-group allows a chiral signature caused by interference between a magnetic dipole and an electronic quadrupole in resonant x-ray Bragg diffraction (intensity enhanced by an electric dipole - electric dipole (E1-E1) absorption event). X-ray data to hand were gathered at the $L_3$ absorption edge [2]. Experimental studies of other compounds hosting heavy transition ions have encountered very different intensities between $L_2$ and $L_3$ absorption edges. Amplitudes in all four channels of polarization contribute to the chiral signature, and corresponding individual intensities are shown to carry useful information worthy of experimental investigation. Contributions to Bragg spots in magnetic neutron diffraction that violate the F-centring condition include a quadrupole exclusive to mixing of $5d^1$ Re manifolds. Based on current knowledge of heavy transition metal ions in an almost octahedral environment, this result is expected. Physically, the quadrupole hallmarks entanglement of anapole and spatial degrees of freedom. Magnetic intensity picked out in neutron polarization analysis is here calculated for all allowed dipoles and quadrupoles. Quadrupoles in x-ray and neutron diffraction amplitudes are predicted to have different angular symmetries, since the former is purely electronic and the latter purely magnetic.

**ACKNOWLEDGEMENT** Dr Y. Tanaka advised on the feasibility of resonant x-ray Bragg diffraction.

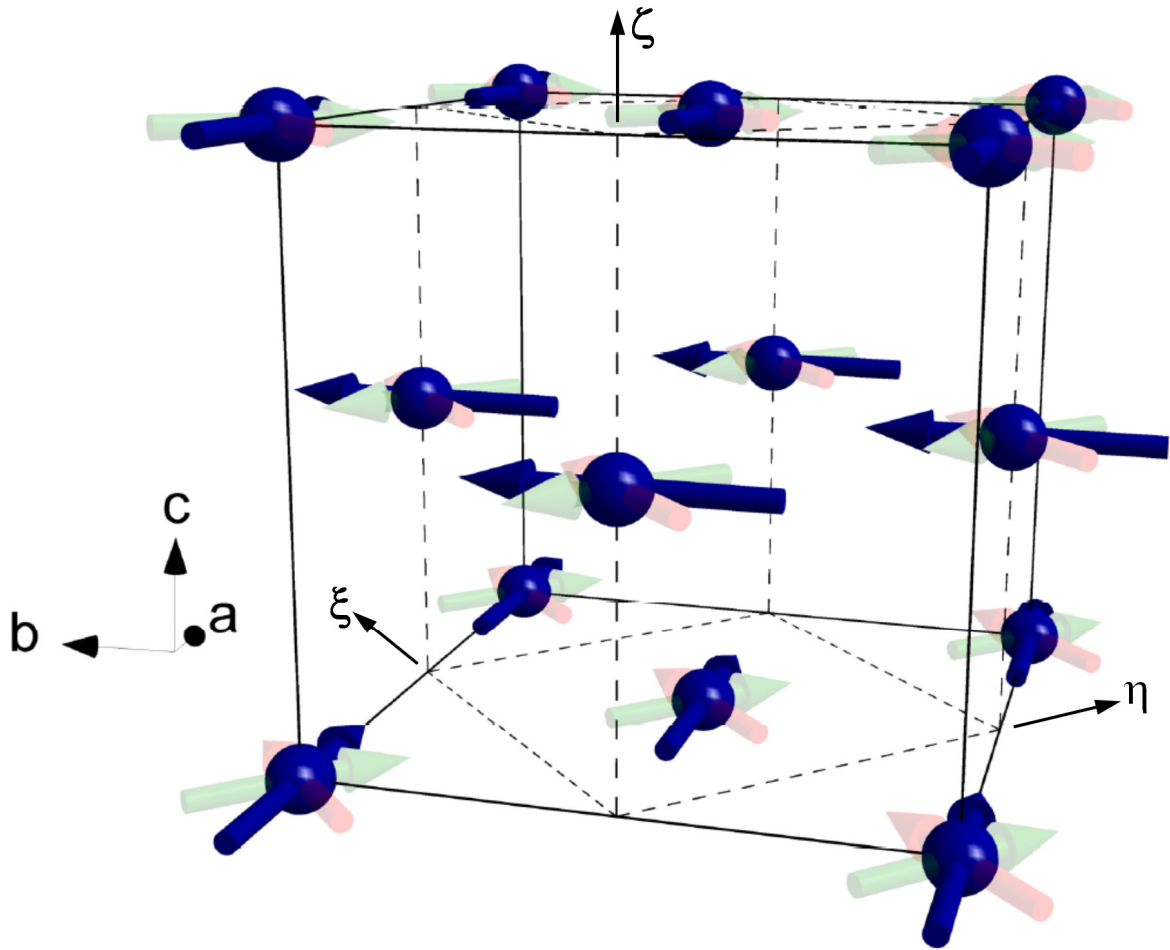

FIG. 1. Ferro- and antiferromagnetic dipole components of the ordered magnetic structure $Ba_2MgReO_6$ are depicted by transparent red and green arrows, respectively. A tetragonal basis labelled ($\xi$, $\eta$, $\zeta$) is introduced in Section II.

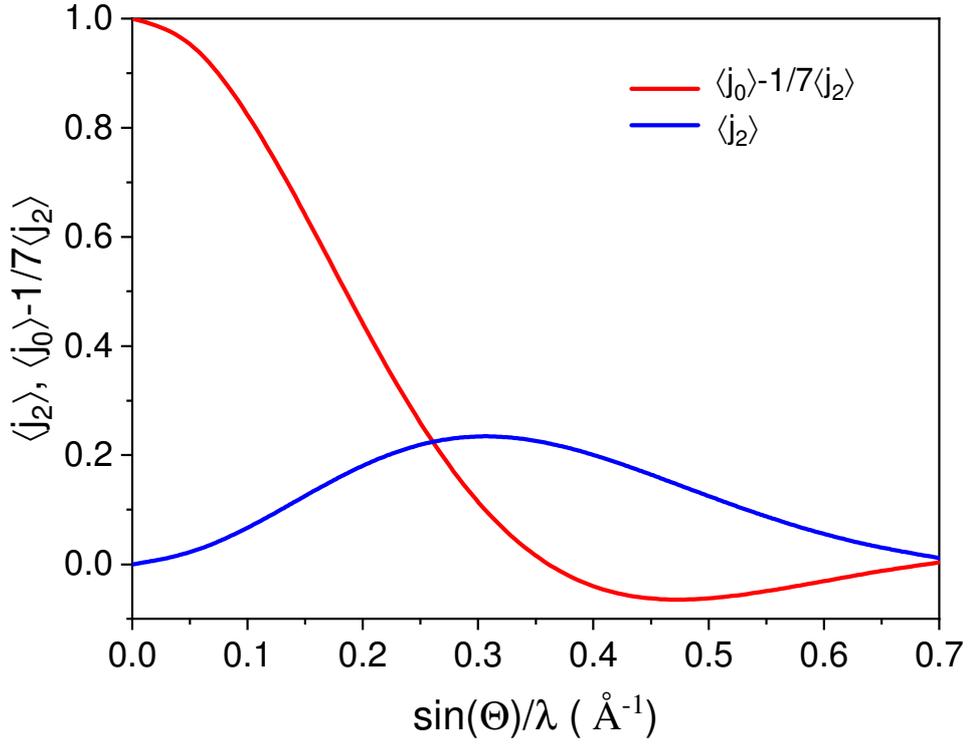

FIG. 2. Radial integrals for $Re^{6+}$ ($5d^1$) as a function of $\sin(\theta)/\lambda = \kappa/4\pi$ in the range 0 - 0.7 Å$^{-1}$ [19]. Blue curve $\langle j_2(\kappa) \rangle$, red curve $[\langle j_0(\kappa) \rangle - (1/7) \langle j_2(\kappa) \rangle]$.

[1] D. Hirai and Z. Hiroi, J. Phys. Soc. Jpn. **88**, 064712 (2019).

[2] D. Hirai, H. Sagayama, S. Gao, H. Ohsumi, G. Chen, Taka-h. Arima, and Z. Hiroi, Phys. Rev. Res. **2**, 022063(R) (2020).

[3] D. D. Maharaj, G. Sala, M. B. Stone, E. Kermarrec, C. Ritter, F. Fauth, C. A. Marjerrison, J. E. Greedan, A. Paramekanti, and B. D. Gaulin, Phys. Rev. Lett. **124**, 087206 (2020).

[4] S. W. Lovesey and D. D. Khalyavin, Phys. Rev. B **102**, 064407 (2020).

[5] G. J. Nilsen, C. M. Thompson, C. Marjerisson, D. I. Badrtdinov, A. A. Tsirlin, and J. E. Greedan, Phys. Rev. B **103**, 104430 (2021).

[6] S. W. Lovesey, D. D. Khalyavin, G. van der Laan, and G. J. Nilsen Phys. Rev. B **103**, 104429 (2021).

[7] S. Gao, D. Hirai, H. Sagayama, H. Ohsumi, Z. Hiroi, and Taka-hisa Arima, Phys. Rev. B **101**, 220412(R) (2020).

[8] S. W. Lovesey, E. Balcar, K. S. Knight, and J. Fernández Rodríguez, Phys. Reports **411**, 233 (2005).

[9] S. W. Lovesey, Phys. Scripta **90**, 108011 (2015).

[10] S. W. Lovesey, *et al*., Phys. Rev. Lett. **122**, 047203 (2019).

[11] D. D. Khalyavin and S. W. Lovesey, Phys. Rev. B **100**, 224415 (2019).

[12] Y. Joly, Y. Tanaka, D. Cabaret, and S. P. Collins, Phys. Rev. B **89**, 224108 (2014).

[13] We use the BNS setting of magnetic space groups, see Bilbao Crystallographic server, http://www.cryst.ehu.es.

[14] B. T. Thole, P. Carra, F. Sette, and G. van der Laan., Phys. Rev. Lett. **68**, 1943 (1992); P. Carra, B. T. Thole, M. Altarelli, and X. D. Wang, *ibid*. **70**, 694 (1993); P. Carra, H. König, B.T. Thole, and M. Altarelli, Physica **192B**, 182 (1993).

[15] L. C. Chapon and S. W. Lovesey, J. Phys.: Condens. Matter **23**, 252201 (2011).

[16] V. Scagnoli and S. W. Lovesey, Phys. Rev. B **79**, 035111 (2009).

[17] S. W. Lovesey and V. Scagnoli, J. Phys.: Condens. Matter **21**, 474214 (2009).

[18] A. Rodríguez-Fernández, J. A. Blanco, S. W. Lovesey, V. Scagnoli, U. Staub, H. C. Walker, D. K. Shukla, and J. Strempfer, Phys. Rev. B **88**, 094437 (2013).

[19] K. Kobayashi, T. Nagao, and M. Ito, Acta. Cryst. A **67**, 473 (2011).

[20] R. M. Moon, T. Riste, and W. C. Koehler, Phys. Rev. **181**, 920 (1969).

[21] J. Jeong, B. Lenz, A. Gukasov, X. Fabreges, A. Sazonov, V. Hutanu, A. Louat, D. Bounoua, C. Martins, S. Biermann, V.Brouet, Y. Sidis, and P. Bourges, Phys. Rev. Lett. **125**, 097202 (2020).